\documentclass[11pt]{article}%

\usepackage{amssymb}
\usepackage{amsmath}
\usepackage{graphics}
\usepackage{epsfig}
\usepackage{amsfonts}
\usepackage{graphicx}%

\usepackage{subfig}

\usepackage{booktabs}

\newcommand{\eqb}{\begin{equation}}
\newcommand{\eqe}{\end{equation}}
\newcommand{\dmb}{\begin{displaymath}}
\newcommand{\dme}{\end{displaymath}}

\newcommand{\eab}{\begin{eqnarray}}
\newcommand{\eae}{\end{eqnarray}}

\newcommand{\e}{\mbox{e}}
\newcommand{\be}{\begin{equation}}
\newcommand{\ee}{\end{equation}}

\begin{document}
\begin{titlepage}
\begin{flushright}
\end{flushright}
\vspace{0.6cm}
\begin{center}
\large{Cosmic Microwave Background as a thermal gas of SU(2) photons:\\ 
Implications for the high-z cosmological model and the value of $H_0$}
\end{center}
\vspace{0.5cm}
\begin{center}\large{Steffen Hahn}
\end{center}
\vspace{0.1cm}
\begin{center}
{\em Karlsruhe Institute of Technology (KIT), Germany;\\ steffen.t.hahn@gmail.com}
\end{center}
\vspace{1.0cm}
\begin{center}
\large{Ralf Hofmann}
\end{center}
\vspace{0.1cm}
\begin{center}
{\em Institut f\"ur Theoretische Physik, Universit\"at Heidelberg,\\ 
Philosophenweg 16, D-69120 Heidelberg, Germany;\\ r.hofmann@thphys.uni-heidelberg.de
}
\end{center}

\begin{abstract}
Presently, we are facing a 3$\sigma$ tension in the most basic cosmological parameter -- the Hubble constant $H_0$. This tension arises when fitting the Lambda-cold-dark-matter model ($\Lambda$CDM) to the high-precision temperature-temperature (TT) power spectrum of the Cosmic Microwave Background (CMB) and to local cosmological observations. 
We propose a resolution of this problem by postulating that the thermal photon gas of the CMB obeys an SU(2) rather than U(1) gauge principle, suggesting a high-$z$ cosmological model which is void of dark matter. Observationally, we rely on precise low-frequency intensity measurements in the CMB spectrum and on a recent model independent (low-$z$) extraction of the relation between the comoving sound horizon $r_s$ at the end of the baryon drag epoch and $H_0$ ($r_s H_0 = \text{const}$). 
We point out that the commonly employed condition for baryon-velocity freeze-out is imprecise, judged by a careful inspection of the formal solution to the associated Euler equation. 
As a consequence, the above mentioned 3$\sigma$ tension actually transforms into a 
5$\sigma$ discrepancy. To make contact with successful low-$z$ $\Lambda$CDM cosmology 
we propose an interpolation based on percolated/depercolated vortices of a Planck-scale axion 
condensate. For a first consistency test of such an all-$z$ model we compute the angular scale of the 
sound horizon at photon decoupling.

\end{abstract}
  
\end{titlepage}

\section{Introduction}
Since the pioneering work by Yang and Mills \cite{YM1954} on the definition of a local four-dimensional, classical, and minimal field theory, which is based on the nonabelian gauge group SU(2), much progress has been made in elucidating the role of topologically stabilized and (anti)\-selfdual field configurations in building the nonperturbative ground state and influencing the properties of its excitations \cite{Adler,BellJackiw1969,BanksCasher,tHooft1986,Fujikawa1980-10,Diakonov,Shuryak}. 
In particular, the deconfining phase is subject to a highly accurate thermal ground state estimate \cite{Hofmann2016a,Bischer2017}, being composed of so-called Harrington-Shepard (anti)calorons \cite{Harrington1977}. 
This (cosmologically relevant) ground state invokes both an adjoint Higgs mechanism \cite{Anderson,Higgs,Guralnik,Englert}, rendering two out of three directions of the SU(2) algebra massive (free thermal quasiparticles), and a U(1)$_\text{A}$ chiral anomaly \cite{Adler,BellJackiw1969,tHooft1986,Fujikawa1980-10}, giving mass to the Goldstone mode induced by the associated dynamical breaking of this global symmetry. 
Radiative corrections to thermodynamical quantities, evaluated on the level of free thermal (quasi)particles, are minute and well under control \cite{Hofmann2016a,Bischer2017}. Note that this is in contrast to the large effects of radiative corrections 
attributed to the effective QCD action at zero temperature in \cite{Pagels,Saviddy} which are exploited as potential inducers of vacuum energy in the cosmological context in \cite{Zhang,PasechnikI,PasechnikII,PasechnikIII,Dona}. However, it was argued in \cite{Casher,Brodsky} that QCD condensates, which contribute to the trace anomaly of the energy-momentum tensor (as implied by the effective action), do not act cosmologically. 

\begin{figure}
\centering
\includegraphics{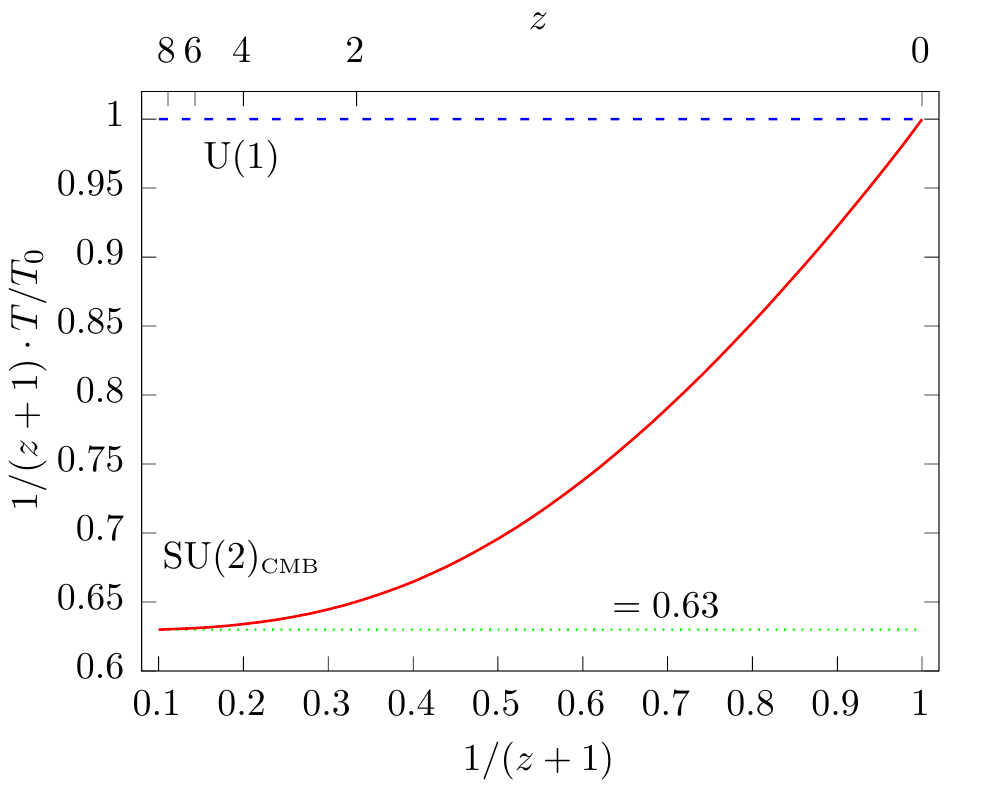}
\caption{\protect{\label{Fig-1}} The $T-z$ scaling relation ${T/(T_0(z+1))}$ in SU(2)$_{\text{CMB}}$ (solid). 
Note the emergence of ${T/T_0}=0.63(z+1)$ for $z\gtrsim 9$ (dotted). 
The conventional U(1) theory for thermal photon gases associates with 
the dashed line. Data taken from \protect\cite{Hofmann2015} after 
slight and inessential correction.}      
\end{figure}

Postulating that thermal photon gases obey an SU(2) rather than a U(1) gauge principle, the SU(2) Yang-Mills scale can be inferred from low-(radio)frequency spectral intensity measurements, e.g. \cite{Arcade2}, of the Cosmic Microwave Background (CMB) \cite{Hofmann2009}, prompting the name SU(2)$_{\text{CMB}}$. 
Below we will use the name SU(2)$_{\text{CMB}}$ synonymously for the implied cosmological model. 
To investigate the consequences of this postulate towards the equation of state radiative corrections are entirely negligible \cite{Hofmann2016a}. 
When subjecting local energy conservation in a Friedmann-Lema\^{\i}tre-Robertson-Walker (FLRW) universe to this equation of state the numerical temperature ($T$) - redshift ($z$) relation ($T(z)$) of the CMB follows, see Fig.\,\ref{Fig-1} \cite{Hofmann2015,HH2017}, where a comparison with the conventional U(1) photon gas is shown. The curvature of ${T/(T_0(z+1))}$ ($T_0=2.725\,$K denoting today's CMB temperature) at low $z$ is due to the influence of the SU(2) Yang-Mills mass scale on the equation of state. 
In \cite{HH2017} an argument is given why recent observational "extractions" of $T(z)$, which claim no deviations from the conventional behavior $T(z)=T_0(z+1)$, are circular.   
One has ${T/T_0}=0.63(z+1)$ at high $z$ and therefore a lower slope compared to the conventional case. 
In an approximation, where recombination at $z_*$ is subjected to thermodynamics, the decoupling condition is $\Gamma_{\text{Th}}(T_*)=H(z_*)$ where $\Gamma_{\text{Th}}$ denotes the Thomson photon-electron scattering rate at the decoupling temperature $T_*\sim 3000\,\mbox{K}$. We have $(\Omega_{0,b}+\Omega_{0,\text{DM}})/\Omega_{0,b}\sim 6.5\equiv R_{m,1}$ where $\Omega_{0,b}$ and $\Omega_{0,\text{DM}}$ denote the respective ratios of today's energy densities in baryons and cold dark matter to the critical density. Since $z_{*,\text{SU}(2)_\text{CMB}}/z_{*,\Lambda\text{CDM}} \sim 1/0.63$ this roughly matches $(1/0.63)^3\sim 4\equiv R_{m,2}$. If a strong matter domination can be assumed during recombination then $R_{m,1}$ should be equal to $R_{m,2}$ but, due to matter-radiation equality occurring at $z\sim 1080$ in SU(2)$_{\text{CMB}}$, this assumption is not quite met, explaining the mild discrepancy between $R_{m,1}$ and $R_{m,2}$. Still, we take this rough argument and the desired minimality of the cosmological model as motivations to omit cold dark matter in the high-$z$ cosmological model which operates down to recombination and well beyond it. 

Concerning the number of massless neutrinos $N_{\nu}$, a conservative input is used: $N_\nu=3$ \cite{Z_0}. This high-$z$ model, composed of SU(2)$_{\text{CMB}}$, baryonic matter, and massless neutrinos ($N_\nu=3$), is sufficient to predict the sound horizon $r_s$ at the end of the baryon drag epoch which, in turn, can be confronted with the $r_s$ - $H_0$ relation, recently extracted from local cosmological observations \cite{Bernal2016}, to determine the value of $H_0$.  
The value of $r_s$, as computed in 
a high-$z$ model, rather sensitively depends on the definition of redshift $z_{\text{drag}}$ for baryon-velocity ($v_b$) freeze-out. 
Usually, $z_{\text{drag}}$ is identified with the maximum position of the so-called drag visibility function $D_\text{drag}$ \cite{Hu1994,Hu1996}. 
However, inspecting the solution $v_b$ of the corresponding Euler 
equation, given as a functional of $D_\text{drag}$, one concludes that this definition applies only in the limit of zero peak width. 
Realistic results for the ionization fraction $\chi_e$, obtained by numerical integration of the according Boltzmann hierarchy (\texttt{recfast} \cite{recfastBarc2015}), imply 
that the width of this peak extends over several hundred units of redshift in both cases $\Lambda$CDM and SU(2)$_{\text{CMB}}$. 
As a consequence, a more precise definition of $z_{\text{drag}}$ is in order which associates with the left flank of $D_\text{drag}$. 
Therefore, we will in the following refer to this corrected redshift for the freeze-out of $v_b$ as $z_{\text{lf, drag}}$. 
Our value $r_s(z_{\text{lf, drag}})\sim 1660$ -- after intersection with the $r_s$ - $H_0$ relation of 
\cite{Bernal2016} -- determines the value of $H_0$ in good agreement with the value obtained in \cite{Riess2016}. 
Also, we would like to point out that, as a consequence of the corrected baryon-velocity freeze-out condition, the value of $H_0$ in $\Lambda$CDM, obtained by this method, is now at a 5$\sigma$ discrepancy with the value published in \cite{Riess2016}.    

To be able to compute the CMB power spectra, our consistent high-$z$ SU(2)$_{\text{CMB}}$ cosmological model 
of Eq.\,(\ref{Hzexp}) needs to be connected to the observationally well cross-checked $\Lambda$CDM low-$z$ parametrization of the Universe's composition. 
To facilitate such an interpolation, a candidate real scalar field $\varphi$ representing the dark sector is the so-called Planck-scale axion (PSA) condensate  \cite{Frieman1995,Giacosa2007,Neubert} which rests on chiral symmetry breaking within the Planckian epoch and the axial anomaly invoked by deconfining thermal ground states of Yang-Mills theories. 
Notice that the only Yang-Mills theory exhibiting the deconfining phase from today to well beyond recombination is SU(2)$_{\text{CMB}}$. 
A model, where $\varphi$ undergoes coherent and damped oscillations at late times such as to effectively represent $\Lambda$CDM, is falsified by the redshift $z_q$, where the Universe's expansion starts to accelerate, being too high. 
This prompts the idea that interpolation between SU(2)$_{\text{CMB}}$ at high $z$ and $\Lambda$CDM at 
low $z$ is achieved by the U(1) topologically stabilized solitonic configurations (vortices) of the 
PSA condensate occurring in percolated form (due to a Berezinskii-Kosterlitz-Thouless phase transition 
following their very creation during a non-thermal phase transition at very high $z$) 
down to intermediate $z$ where a depercolation transition partially liberates them to effectively represent a pressureless vortex gas. Whether or not the cores of depercolated PSA vortices properly serve as dark-matter 
halos in spiral galaxies to explain the observed flattening of 
rotation curves and the lensing signatures of bullet galaxies is an open question. Likewise, it 
is not yet guaranteed that this new cosmological model, which exhibits radiation domination and baryon freeze-out prior to 
photon decoupling, explains the observed angular power spectra of the CMB.   

This work is organized as follows. 
In Sec.\,\ref{CM} we explain our high-$z$ cosmological model SU(2)$_{\text{CMB}}$, introduced in \cite{HH2017}, and compare it with the conventional $\Lambda$CDM cosmology. 
The modification of decoupling conditions due to finite-widths visibility functions is discussed in Sec.\,\ref{RC}. 
Based on this, we perform the computation of $r_s$ and confront it with the $r_s$-$H_0$ relation of \cite{Bernal2016} to determine the value of $H_0$. 
Subsequently, in Sec.\,\ref{PSA} we investigate whether coherent and damped oscillations of the PSA field can realistically represent $\Lambda$CDM at low $z$ -- with a negative result. 
According to \cite{HH2017} we are thus led to propose an interpolation between high-$z$ SU(2)$_{\text{CMB}}$ and low-$z$ $\Lambda$CDM in terms of percolated PSA vortices which, at some intermediate redshift $z_p$, partially undergo a depercolation transition. 
Such a model is demonstrated to be consistent with the extremely well observed angular scale of the sound horizon at photon 
decoupling \cite{PlanckCosParams}. Finally, we summarize our results and provide an outlook on how the new model can be tested further by confrontation with the power spectra of various CMB angular correlation functions.  

\section{Definition of cosmological model SU(2)$_{\text{CMB}}$\label{CM}}
In a flat FLRW universe, a cosmological model is given in terms of the $z$-dependence of the Hubble parameter  
\eqb
\label{Hz}
H(z)=H_0\,\sqrt{\sum_{i}\Omega_i(z)}\,,
\eqe
where $H_0$ is today's cosmological expansion rate and $\Omega_i(z)=f_i(z)  \Omega_{i,0}$.
Here $\Omega_{i,0}$ is the fraction of the energy density $\rho_{i,0}$ of fluid $i$ to the critical density $\rho_{c,0}$ today.
The function $f_i(z)$ is determined by energy conservation subject to fluid $i$'s equation of state.
From now on we work in supernatural units ($c=\hbar=k_\text{B}=1$) where Newton's constant $G$ has units of inverse mass.
Table~\ref{TabcsomPara} lists the parameter values used subsequently.

\subsection{The conventional $\Lambda$CDM model}
In the conventional high-$z$ $\Lambda$CDM model is $H(z)$ given as
\begin{equation}
\label{Hzexpconv}
H(z)=H_0\,\left[\left(\Omega_{b,0}+\Omega_{\text{CDM},0}\right)\,(z + 1)^3 +\left(1+\frac{7}{8}\left(\frac{4}{11}\right)^{4/3}\,N_{\text{eff}}\right)
\Omega_{\gamma,0}(z + 1)^4\right]^{1/2}\,.
\end{equation}
Here non-relativistic matter  decomposes into baryonic ($b$) and cold dark matter (CDM). 
The radiation component contains photons with two polarizations, two relativistic vector modes with three polarizations each, and $N_\text{eff}$ flavors of massless neutrinos with two polarizations each. $\Omega_{\gamma,0}$ is today's fraction of photonic to the critical energy density\footnote{For details see \cite{Ade2016}.}.

\subsection{Modifications of $\Lambda$CDM towards SU(2)$_{\text{CMB}}$}
In high-$z$ SU(2)$_{\text{CMB}}$ the Hubble parameter is given as 
\begin{equation}
H(z)=H_0\,\left[\Omega_{b,0}\,(z + 1)^3 + 4\cdot \left(0.63\right)^4 \left(1+\frac{7}{32}\left(\frac{16}{23}\right)^{4/3}\,N_{\nu}\right)
\Omega_{\gamma,0}(z + 1)^4\right]^{1/2}\,. 
\label{Hzexp}
\end{equation}
In this case, only baryonic matter is present. We reiterate that both models, Eq.\,(\ref{Hzexpconv}) and Eq.\,(\ref{Hzexp}),  
need to be supplemented by a dark sector to yield successful low-$z$ $\Lambda$CDM cosmology, see Eq.\,(\ref{DSed}). The radiation sector is modified due to a different number of relativistic degrees of freedom and due to the SU(2)$_{\text{CMB}}$ high-$z$ temperature-redshift relation $T(z)$, for details see \cite{Hofmann2015, HH2017}.

\begin{table*}
	\centering
	\caption{Cosmological parameter values employed in the computations and their sources, taken from \cite{HH2017}.\label{TabcsomPara}}
	\begin{tabular}{lcl}
	   & & \\
		\toprule
		parameter & value  & source  \\
		\midrule
		$H_0$ (SU$(2)_{\text{CMB}}$) & $(73.24\pm 1.74)$\,km\,s$^{-1}$\,Mpc$^{-1}$  & \cite{Riess2016}  \\
		$H_0$ ($\Lambda$CDM) & $(67.31\pm 0.96)$\,km\,s$^{-1}$\,Mpc$^{-1}$  & TT+lowP, \cite{Ade2016}   \\
		$T_0$ & 2.725\, K &  \cite{Fixsen1996}\\
		$\Omega_{\gamma,0} h^2$ & $2.46796\times 10^{-5}$.  & based on $T_0=2.725\,$K \\
		$\Omega_{b,0} h^2$ & $0.02222 \pm 0.99923$ & TT+lowP \cite{Ade2016} \\
		$\Omega_{\text{CDM},0} h^2$ & $0.1197\pm 0.0022$& TT+lowP, \cite{Ade2016} \\
		$\eta_{10}$ & $6.08232\pm 0.06296$ & based on $\Omega_{\gamma,0}h^2$, TT+lowP \cite{Ade2016}  \\
		$Y_P$ & $0.252\pm 0.041$ & TT, \cite{Ade2016}\\ 
		$N_{\text{eff}}$ & $3.15\pm 0.23$ & abstract, \cite{Ade2016} \\ 
		\bottomrule
	\end{tabular}
\end{table*}

\section{The end of recombination\label{RC}}
The comoving sound horizon $r_s$ at redshift $z$ is defined as
\eqb
\label{shdef}
r_s(z)=\int_z^\infty dz^\prime\,\frac{c_s(z^\prime)}{H(z^\prime)}\,,
\eqe
whereby $c_s$ is the sound velocity in the primordial baryon-electron-photon plasma, given as
\eqb
\label{soundvel}
c_s\equiv\frac{1}{\sqrt{3(1+R)}}\,.
\eqe
The function $R(z)$ is determined by 3/4 of the ratio of energy densities in baryons and photons. 
In $\Lambda$CDM we have
\eqb
\label{RofetaLambda} 
R(z)\equiv 111.019\,\frac{\eta_{10}}{z+1}\,,
\eqe
whereas in SU(2)$_{\text{CMB}}$ one obtains  
\eqb
\label{RofetaZ} 
R(z)\equiv 111.019\,\frac{\eta_{10}}{(0.63)^4(z+1)}\,. 
\eqe
The values of $\eta_{10}$ can be read off Table~\ref{TabcsomPara}.

\subsection{Conventional freeze-out}
The final stages of recombination can be characterized in a twofold way.
One considers either (i) photon temperature freeze-out, which is relevant for the peak structure in the temperature-temperature (TT) angular power spectrum of the CMB, or (ii) baryon velocity freeze-out, which is detectable in the matter correlation function (galaxy counts).
Concerning case (i), the conventional criterion, which fixes the redshift $z_*$, reads
\eqb
\label{optdepthThomson}
\tau(z_*)=\sigma_T\int_0^{z_*}dz\,\frac{\chi_e(z)n_e^b(z)}{(z+1)H(z)}=1\,,
\eqe
where $\sigma_T$ denotes the total cross section for Thomson scattering, $\chi_e$ is the ionization fraction (calculated with \texttt{recfast}), and $n_e^b$ refers to the density of free electrons just before hydrogen recombination, given as
\begin{equation}
n_e^b(z) = 410.48\cdot10^{-10}\,\eta_{10}(1-Y_P)(z+1)^3\,\mbox{cm}^{-3} \label{neb}\,.
\end{equation}
Here $Y_P$ denotes the helium mass fraction in baryons (see Table~\ref{TabcsomPara}).
Concerning case (ii), the conventional criterion is defined as
\eqb
\label{dragdepth}
\tau_{\rm drag}(z)=\sigma_T\int_0^{z}dz^\prime\,\frac{\chi_e(z^\prime)n_e^b(z^\prime)}{(z^\prime+1)H(z^\prime)R(z^\prime)} = 1\,.
\eqe 

\subsection{Corrected freeze-out}
\begin{figure}
\centering
\subfloat{\includegraphics{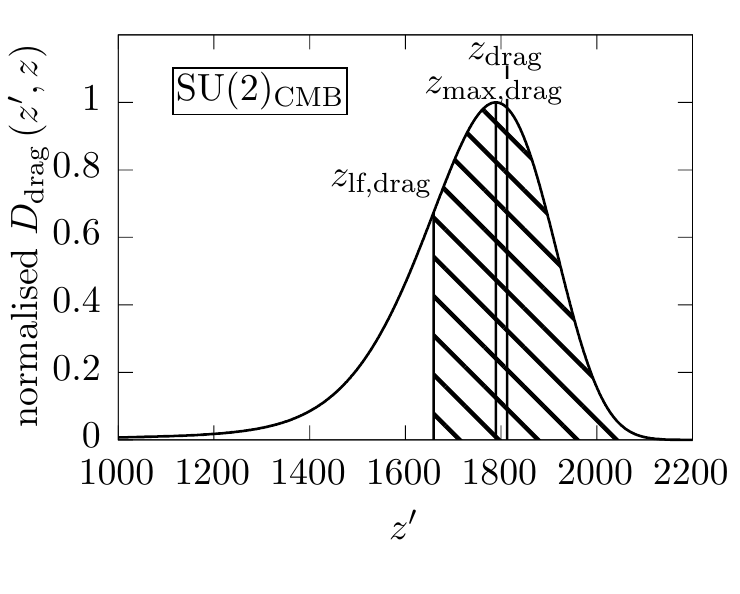}}
\subfloat{\includegraphics{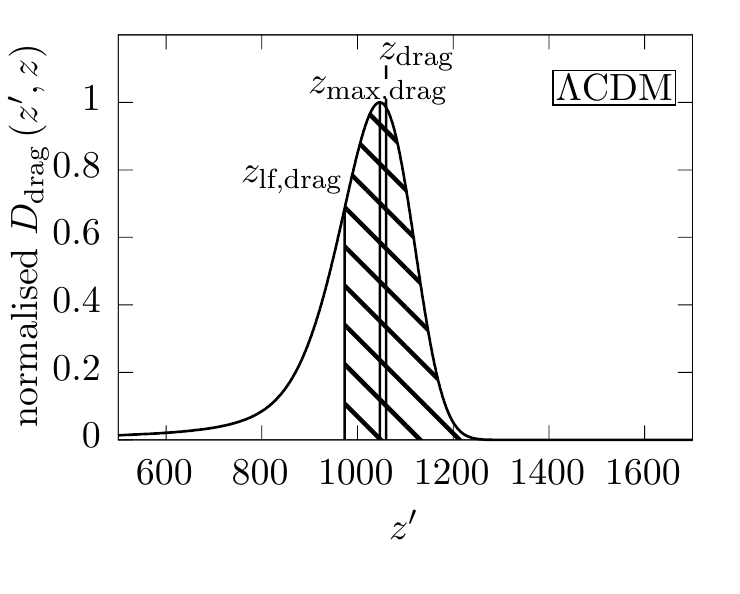}}
\caption{Normalised function $D_{\rm drag}(z^\prime,z)$, defined in Eq.\,(\ref{Ddef}), 
if $z\le z_{\rm max,drag}$ for SU(2)$_{\rm CMB}$ (left) and $\Lambda$CDM (right). Redshift $z_{\rm lf,drag}$ is defined 
as the position of the maximum of $\frac{dD_{\rm drag}}{dz^\prime}$ (position of left flank of 
$D_{\rm drag}$) whereas $z_{\rm max,drag}$ denotes the position of the maximum of 
$D_{\rm drag}$. The value of $z_{\rm drag}$, defined in Eq.\,(\ref{dragdepth}), essentially 
coincides with $z_{\rm max,drag}$: $z_{\rm drag}=1813\sim z_{\rm max,drag}=1789$ for SU(2)$_{\rm CMB}$ 
and $z_{\rm drag}=1059\sim z_{\rm max,drag}=1046$ for $\Lambda$CDM. This should be contrasted 
with $z_{\rm lf,drag}=1659$ for SU(2)$_{\rm CMB}$ and $z_{\rm lf,drag}=973$ for $\Lambda$CDM. The hatched area under the curve determines the freeze-out value of $v_b/(z+1)$. \protect{\label{Fig-C1}}.}     
\end{figure}

We now show that conditions (\ref{optdepthThomson}) and (\ref{dragdepth}) are imprecise due to the finite widths of the respective visibility functions.
To see this, we have to analyze the formal solution of the Boltzmann hierarchy for the temperature perturbation and of the Euler equation for $v_b$ \cite{Hu1994,Hu1996,BondEfstathiou1984}.
Since the argument is similar for both cases we focus on the latter only.
The Euler equation reads
\eqb
\label{Euler}
\dot{v}_b=\frac{\dot{z}}{z+1}v_b+k\Psi+\dot{\tau}_{\rm drag}(\Theta_1-v_b)\,, 
\eqe
where $k$ is the comoving wave number (omitted as a subscript in the following), $\Theta_1$ denotes the (relative) dipole of the temperature anisotropy \cite{PeeblesWilk}, and $\Psi$ represents the Newtonian gravitational potential.
The overdot demands differentiation with respect to conformal time.
Transforming the conformal time to a redshift dependence, the solution of Eq.~(\ref{Euler}) is 
\begin{align}
\label{FullEulerz}
\frac{v_b(z)}{z+1}&=\lim_{Z\nearrow \infty}\int_{z}^{Z} dz^\prime\,\frac{\e^{-\tau_{\rm drag}(z^\prime,z)}}{H(z^\prime) (z^\prime+1)}\left(\dot{\tau}_{\rm drag}(z^\prime)\Theta_1(z^\prime)+k\Psi(z^\prime)\right)\nonumber\\ 
&\sim\lim_{Z\nearrow \infty}\int_{z}^{Z} dz^\prime\,D_{\rm drag}(z^\prime,z)\Theta_1(z^\prime)\,.
\end{align}
Here $\tau_{\rm drag}$ is defined as
\eqb
\label{tauz}
\tau_{\rm drag}(z^\prime,z)\equiv \int_{z}^{z^\prime}dz^{\prime\prime}\,\frac{\dot{\tau}_{\rm drag}(z^{\prime\prime})}{H(z^{\prime\prime})}\,,
\eqe
and the visibility function $D_{\rm drag}(z^\prime,z)$ is represented by
\eqb
\label{Ddef}
D_{\rm drag}(z^\prime,z)\equiv \frac{\e^{-\tau_{\rm drag}(z^\prime,z)}
\dot{\tau}_{\rm drag}(z^\prime)}{H(z^\prime) (z^\prime+1)}\,.
\eqe 
In order to study freeze-out the function $\Theta_1$ in Eq.~(\ref{FullEulerz}) is considered slowly varying.
Therefore, the variability integral solely depends on $D_\text{drag}$ within its peak region.
In both cases $\Lambda$CDM and SU(2)$_{\rm CMB}$ function $D_\text{drag}$ exhibits a broad peak in dependence of $z^\prime$ whose shape and maxima does not depend on $z$, see Fig.~\ref{Fig-C1}. 
Note that Eq.~(\ref{dragdepth}) describes the maxima $z_{{\rm max},\text{drag}}^\prime$ of $D_{\text{drag}}(z^\prime,z)$.
However, due to the finite width the integral in Eq.~(\ref{FullEulerz}) is not saturated at $z=z_{{\rm max},\text{drag}}$ but rather ceases to vary for $z<z_{\text{lf},\text{drag}}$ where lf denotes the maxima of the $z^\prime$ derivative of $D_\text{drag}$.
Therefore, $z_{\text{lf},\text{drag}}$ defines the freeze-out point more realistically than $z_{{\rm max},\text{drag}}$.
According to Fig.~\ref{Fig-C1}'s caption the values of $z_{\rm drag}, z_{\text{lf},\text{drag}}$ deviate substantially. 
Namely,

\begin{align}
z_{\text{drag}}=1813, \quad z_{{\rm max},\text{drag}} &= 1789, \quad z_{{\rm lf},\text{drag}} = 1659 \quad \left( \text{SU(2)}_{\rm CMB} \right)\,, \nonumber \\
z_{\text{drag}}=1059, \quad z_{{\rm max},\text{drag}} &= 1046, \quad z_{{\rm lf},\text{drag}} = \phantom{0}973 \quad \left( \Lambda\text{CDM} \right) \, \label{eq:dragShifts}.
\end{align}
An analogous discussion applies to photon temperature freeze-out with the following results (see \cite{HH2017}): 
\begin{align}
z_{*}=1694, \quad z_{{\rm max},*} &= 1694, \quad z_{{\rm lf},*} = 1555 \quad \left( \text{SU(2)}_{\rm CMB} \right)\,, \nonumber \\
z_{*}=1090, \quad z_{{\rm max},*} &= 1072, \quad z_{{\rm lf},*} = \phantom{0}988 \quad \left( \Lambda\text{CDM} \right) \, \label{eq:starShifts} .
\end{align}

\section{The value of $H_0$\label{H0}}
\begin{figure}
\centering
\includegraphics{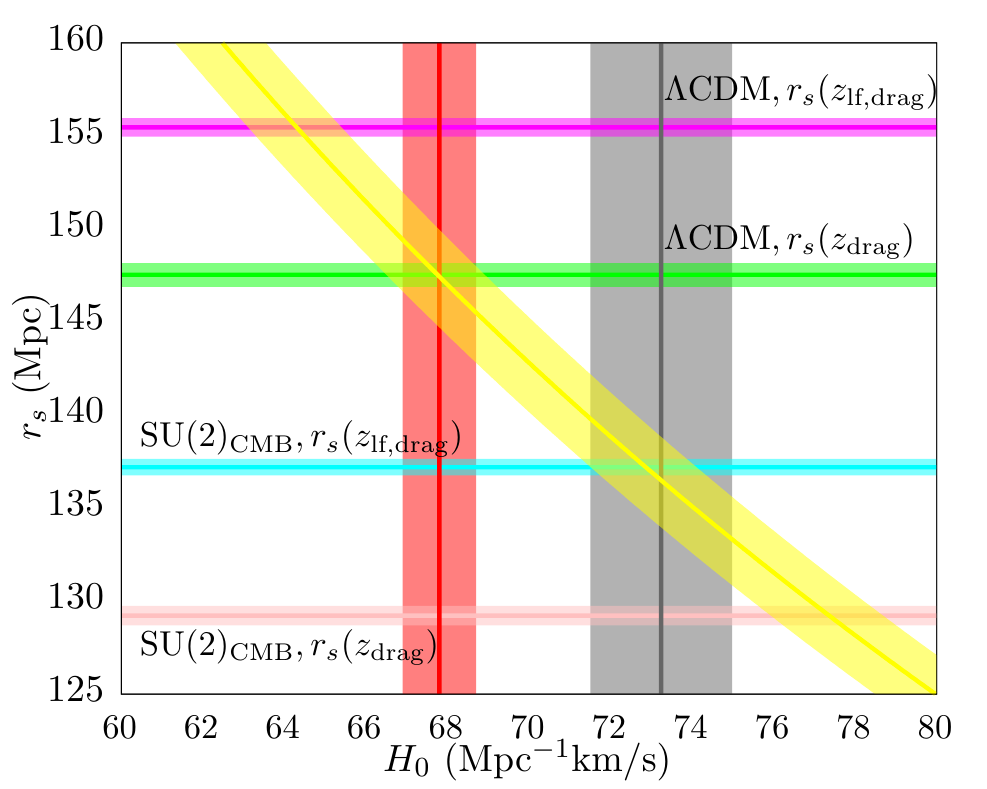}
\caption{\protect{\label{fig:confrontation}} The $r_s(z_{\rm lf, drag})$-$H_0$ relation (curved band) of \cite{Bernal2016} in confrontation with the high-$z$ predictions of $r_s(z_{\rm lf, drag})$ and $r_s(z_{\rm drag})$ in $\Lambda$CDM and SU(2)$_{\rm CMB}$ (horizontal bands) of Eqs.\,(\ref{eq:rsDragValues}). Vertical bands indicate the values of $H_0$ extracted in \cite{Ade2016} (low) and in \cite{Riess2016} (high). Note that  there is a $\sim$3$\sigma$ tension. However, a $\sim$7$\sigma$ discrepancy exists between the $H_0$ values of $(64.3\pm 1.1)$\,km\,s$^{-1}$\,Mpc$^{-1}$ and $(72.9\pm 1.2)$\,km\,s$^{-1}$\,Mpc$^{-1}$ associated with the intersections of $r_s(z_{\rm lf, drag})$ in
$\Lambda$CDM and in SU(2)$_{\rm CMB}$, respectively, with the $r_s(z_{\rm lf, drag})$-$H_0$ relation. Taking $H_0=(73.24\pm 1.7)$\,km\,s$^{-1}$\,Mpc$^{-1}$ from \cite{Riess2016} the discrepancy between this value and $(64.3\pm 1.1)$\,km\,s$^{-1}$\,Mpc$^{-1}$ is about 5$\sigma$. }      
\end{figure}

Subjecting the freeze-out redshifts of (\ref{eq:dragShifts}) to Eqs.~(\ref{shdef}) under consideration of Eqs.~(\ref{Hzexpconv}) and ~(\ref{Hzexp}), yields
\begin{align}
r_s(z_{\rm drag})&=(129.22\pm 0.52)\,\mbox{Mpc}\ \ \ \ (\mbox{SU(2)}_{\text{CMB}})\,,\nonumber\\ 
r_s(z_{\rm lf,drag})&=(137.19\pm 0.45)\,\mbox{Mpc}\ \ \ \ (\mbox{SU(2)}_{\text{CMB}})\,,\nonumber\\ 
r_s(z_{\rm drag})&=(147.33\pm 0.49)\,\mbox{Mpc}\ \ \ \ (\Lambda\mbox{CDM})\,,\nonumber\\ 
r_s(z_{\rm lf,drag})&=(154.57\pm 3.33)\,\mbox{Mpc}\ \ \ \ (\Lambda\mbox{CDM})\, \label{eq:rsDragValues}.
\end{align}
In Fig.~\ref{fig:confrontation}, these ($H_0$ independent) values of the sound horizon are confronted with the $r_s$-$H_0$ relation of \cite{Bernal2016}. 
Note the good agreement between the values of $H_0$ implied by $r_s(z_{\rm lf,drag})$ in SU(2)$_{\text{CMB}}$ and the extraction performed in \cite{Riess2016}.
On the other hand, $r_s(z_{\rm drag})$ reproduces the value of $H_0$ published in \cite{Ade2016} which exhibits a 3$\sigma$ tension compared to \cite{Riess2016}.
However, according to Fig.\,\ref{fig:confrontation}, the more realistic freeze-out value $z_{\rm lf,drag}$ in $\Lambda$CDM entails 
\begin{equation}
H_0= (64.5\pm 1)\,\text{km}\,\text{ s}^{-1}\,\text{Mpc}^{-1} \,.
\end{equation}
Thus, in $\Lambda$CDM there actually is a 5$\sigma$ discrepancy between the value of $H_0$ quoted in \cite{Riess2016} and obtained by confrontation of $r_s$ with the $r_s$-$H_0$ relation of \cite{Bernal2016}.

\section{Planck-scale-axion field and interpolation of high-$z$ with low-$z$ cosmology\label{PSA}}

Here we would like to analyze cosmological models which link low-$z$ $\Lambda$CDM with high-$z$ SU(2)$_{\rm CMB}$. 
We assume a dark sector which originates from a real, minimally coupled scalar field -- a pseudo Nambu-Goldstone mode of dynamical chiral symmetry occurring at the Planck scale \cite{Frieman1995,Giacosa2007} -- whose potential is due to the chiral U$(1)_\text{A}$ anomaly invoked by (anti)calorons of the deconfining, thermal ground state of Yang-Mills theories 
\cite{YM1954,Adler,BellJackiw1969,Fujikawa1980-10,Fujikawa1979,Peccei1977a,Peccei1977b}. 
This prompts the name Planck-scale axion (PSA). The only Yang-Mills theory, which is deconfining well above recombination, is SU(2)$_{\rm CMB}$ because otherwise there wouldn't be just one species of photons. 

The radiatively protected potential for the axion condensate $\varphi$, arising due to the thermal 
ground state of SU(2)$_{\rm CMB}$ \cite{Peccei1977a,Peccei1977b}, reads as follows 
\begin{equation}
\label{potxaion}
V\left(\varphi \right) = \left(\kappa \Lambda_\text{CMB}\right)^4 \cdot \left(1-\cos\left(\varphi/m_\text{P}\right)\right)\,,
\end{equation}
where $\Lambda_{\rm CMB}\sim 10^{-4}\,$eV, $\kappa$ is a dimensionless factor of order unity, and the reduced Planck mass reads 
\eqb
\label{defplamass}
m_{\rm P}\equiv \frac{1.22\times 10^{19}}{\sqrt{8\pi}}\,\mbox{GeV}=(8\pi G)^{-1/2}\,.
\eqe
With a canonical kinetic term for $\varphi$ the according equation of motion is 
\begin{equation}
\label{eomaxion}
\ddot\varphi + 3 H \dot\varphi + \frac{\mathrm d}{\mathrm d \varphi} V\left(\varphi\right) = 0\,,
\end{equation}
where an overdot signals the derivative with respect to cosmological time.

In a first attempt at a $\Lambda$CDM - SU(2)$_{\rm CMB}$ interpolation we assume spatially homogeneous $\varphi$-field 
dynamics subject to $\Lambda$CDM constraints at low $z$. It turns out, however, that such a model predicts a value of $z_q$, defined as the zero of the deacceleration parameter 
\eqb
\label{defdec}
q(z)\equiv\frac{z+1}{2 \hat H^2} (\hat H^2)' - 1\,,
\end{equation}
of about $z_q\sim 3$ which is much higher than the realistic value $\sim 0.7$ 
obtained in $\Lambda$CDM. Therefore, as a second proposal we abolish the energy density arising from {\sl spatially homogeneous} configurations 
of the field $\varphi$. Rather, we conceive the dark-matter sector in $\Lambda$CDM as a piece of energy density 
due to depercolated topological solitons (vortices) of the field $\varphi$ which percolate instantaneously into a dark-energy like piece at some redshift $z_p$ such that $z_{\rm re}\ll z_p\ll z_{\rm lf, drag}$. 
The origin of such a vortex percolate, with hierarchically ordered core sizes, could be due to Hagedorn transitions of Yang-Mills theories in the early universe which are accompanied by Berezinskii-Kosterlitz-Thouless transitions in the axionic sector. 
Today's value of $\Omega_{\Lambda}$ would then be interpreted in terms of not-yet depercolated vortices. 
Indeed, in such an interpolation between $\Lambda$CDM and SU(2)$_{\rm CMB}$ a value of $z_p\sim 155$ can be fitted to the angular size of the sound horizon at photon decoupling. 
At $z_{\rm lf, drag}$ the extra contribution to dark-energy amounts to $\sim 0.65\%$ of the baryonic energy density which is consistent with SU(2)$_{\rm CMB}$.       

\subsection{Spatially homogeneous, coherent oscillations}

Here we discuss a cosmological model where the interpolation between $\Lambda$CDM and 
SU(2)$_{\rm CMB}$ is attempted by a spatially homogeneous PSA field which undergoing damped and coherent oscillations at late times. 
This models a pressureless component (cold dark matter) and component with negative pressure (dark energy). 
Notice that these two components represent fluids that are not separately conserved. 

\begin{figure}
\centering
\includegraphics{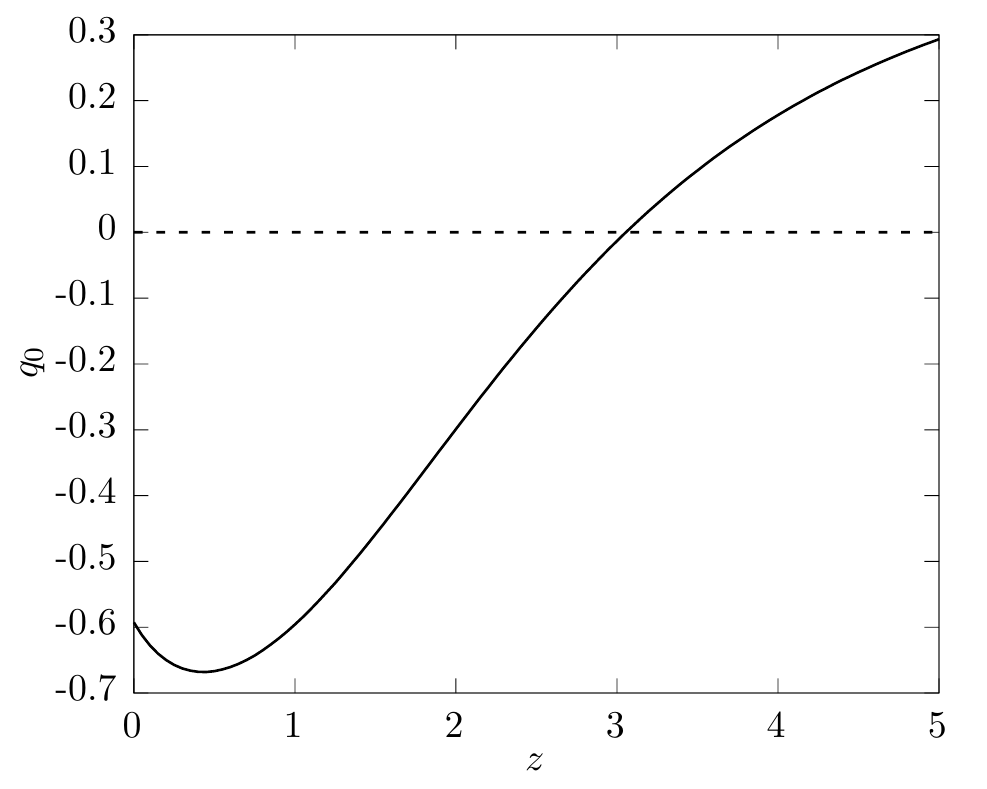}
\caption{\protect{\label{Fig-3}}The deacceleration parameter $q(z)$ of Eq.\,(\ref{defdec}) for the model defined by Eqs.\,(\ref{hattedphi}), (\ref{hattedH}), (\ref{c1}), and (\ref{c2}). Notice that the value of the zero $z_q$ of $q(z)$ is $z_q\sim 3$. This is much higher than 
the realistic value $z_q\sim 0.7$ obtained in $\Lambda$CDM.}      
\end{figure}

The Hubble equation reads 
\begin{equation}
\label{Haxion}
H^2 = \frac{8 \pi G}{3} \left(\frac{1}{2} \dot\varphi^2 + V\left(\varphi\right) + \rho_b + \rho_r \right)
\equiv \frac{8 \pi G}{3}\rho_c\,.
\end{equation}
Here $\rho_r$ denotes radiation-like energy density including SU(2)$_{\rm CMB}$ (for $z\le 9$ radiation energy density is severely suppressed in the cosmological model, for $z>9$ the thermal ground state and the masses of the vector modes of SU(2)$_{\rm CMB}$ can be neglected) and three flavours of massless neutrinos, $\rho_b$ is the energy density of baryons, in addition to the energy density $\frac{1}{2} \dot\varphi^2 + V\left(\varphi\right)$ associated with the spatially homogeneous PSA field $\varphi$ which evolves temporally. 
Eqs.\,(\ref{eomaxion}) and (\ref{Haxion}) can be cast into fully dimensionless equations by rescaling with powers of $m_{\rm P}$ in the following way 
\begin{equation}
V = m_\text{P}^4 \hat V\,,\ \ \  \rho_{i} = m_\text{P}^4\hat \rho_{i}\, \, (i=b,r)\,,\ \ \  \varphi = m_\text{P} \hat{\varphi}\,,\ \ \  
H = m_\text{P} \hat H\,.
\end{equation}
In general, dimensionless quantities (after rescaling with the appropriate power of $m_{\rm P}$) are indicated by the hat-symbol. 
After rescaling and in dependence of $z$ Eqs.\,(\ref{eomaxion}) and (\ref{Haxion}) transmute into
\eqb
\label{hattedphi}
\hat\varphi'' \left[\left(z+1\right) \hat H\right]^2 + \hat \varphi'\left[\frac12\left(z+1\right)^2
(\hat H^2)' - 2\left(z+1\right) \hat H^2\right] + \widehat{\frac{\mathrm dV}{\mathrm d \varphi}}=0 
\eqe
and 
\eqb
\label{hattedH}
\hat H^2 = \frac{1}{3} \frac{\hat V+\hat \rho_{b,0} \left(z+1\right)^3+\hat\rho_{r} }{1-\frac{1}{6} \left(z+1\right)^2 \hat\varphi'^2}\,, 
\eqe
where a prime demands $z$-differentiation. 
In Eq.\,(\ref{hattedH}) we approximate $\hat\rho_r$ as  
\begin{equation}
\label{radne}
\hat\rho_r = \hat \rho_{\gamma,0} \cdot \left\{ \begin{array}{cl}0 &\quad  (z < 9)  \\ 
4(0.63)^3 \left(1+\frac{7}{32}\left(\frac{16}{23}\right)^{4/3}\, 3\right)\left(z+1\right)^4 & \quad (z \geq 9)\,. 
\end{array}\right. 
\end{equation}
With the initial conditions 
\eqb
\label{ic}
\hat{\varphi}(z=z_i)=\hat{\varphi}_i\,,\ \ \ \  \hat{\varphi}'(z=z_i)=0
\eqe
for sufficiently large $z_i$ (no roll; in practice one safely can chose $z_i\sim 50$) the solution to Eq.\,(\ref{hattedphi}) subject to Eq.\,(\ref{hattedH}) is unique. 
To fix the values of $\kappa$ in Eq.\,(\ref{potxaion}) and $\hat{\varphi}_i$ in Eq.~(\ref{ic}) we demand 
\eqb 
\label{c1}
\rho_{c,0}=\frac{3 H_0^2}{8 \pi G} =3\,m_\text{P}^4 \hat{H}_0^2\,
\eqe
and that $\Omega_{\Lambda}$ coincides with typical fit value $\Omega_{\Lambda}\sim 0.7$ obtained in $\Lambda$CDM cosmology 
\cite{PlanckCosParams}:
\begin{equation}
\label{c2}
\Omega_{\Lambda}=\frac{m_\text{P}^4}{\rho_{c,0}}\lim_{z\searrow 0}\left(\hat{V} - \frac{1}{2} \left(\left(z+1\right) \hat H \hat\varphi'\right)^2\right)\sim 0.7\,.
\end{equation}
Fig.\,\ref{Fig-3} shows the deacceleration parameter $q(z)$ for the model defined by Eqs.\,(\ref{hattedphi}), (\ref{hattedH}), (\ref{c1}), and (\ref{c2}). 
Obviously, this model is falsified by a much too high value of the zero $z_q$ of $q(z)$.  

\subsection{Percolated and unpercolated vortices}

Here the basic idea invokes the fact that a PSA field $\varphi$, due to non-thermal phase transitions of the Hagedorn type (e.g., there should be an SU(2)$_e$ Yang-Mills theory of scale $\Lambda_e\sim 0.5\,$MeV going confining at $T\sim\Lambda_e$) is subject to U$(1)_\text{A}$ winding and in this way creation of a density of percolated topological solitons (vortex percolate) with a hierarchical ordering of core sizes. 
Percolation could be understood as a Berezinskii-Kosterlitz-Thouless phase transition \cite{Bere1972,KT1973}. 
Effectively, this percolate represents homogeneous, constant energy density. 
As the universe expands the vortex percolate is increasingly stretched, and, at around some critical redshift $z_p\ll z_{\rm lf, drag}$ it releases a part of its solitons characterized by some specific core size. 
The ensuing vortex gas acts cosmologically like pressureless matter. 
Vortices of larger core sizes remain trapped in the percolate. 
For this scenario to be a consistent interpolation of SU(2)$_{\rm CMB}$ and $\Lambda$CDM we need to assure that $z_p\gg z_{\rm re}\sim 6$ \cite{QuasarGunnPeterson}.    

With the definition of Eq.\,(\ref{radne}) the cosmological model to be considered thus reads
\begin{equation}
\label{Haxion2}
\hat H^2 = \frac{1}{3} \left(\hat\rho_b + \hat\rho_{\rm DS} + \hat\rho_r \right)\,,
\end{equation}
where $\rho_{\rm DS}$ is the dark-sector energy density, defined as
\begin{equation}
\label{DSed} 
\hat\rho_{\rm DS }=\hat{\rho}_\Lambda+\hat{\rho}_{\text{CDM},0} \cdot \left\{ \begin{array}{cl}
\left(z_{\phantom{p}}+1\right)^3 & \quad (z<z_p) \\
\left(z_p+1\right)^3 & \quad (z\geq z_p)\,, 
\end{array}  \right.
\end{equation}
where $\hat{\rho}_\Lambda$ and $\hat{\rho}_{\text{CDM},0}$ are today's values of the dark-energy and cold-dark-matter densities associated with Eq.\,(\ref{c2}) and the value quoted in Table 1, respectively. 

\begin{figure}
\centering
\includegraphics{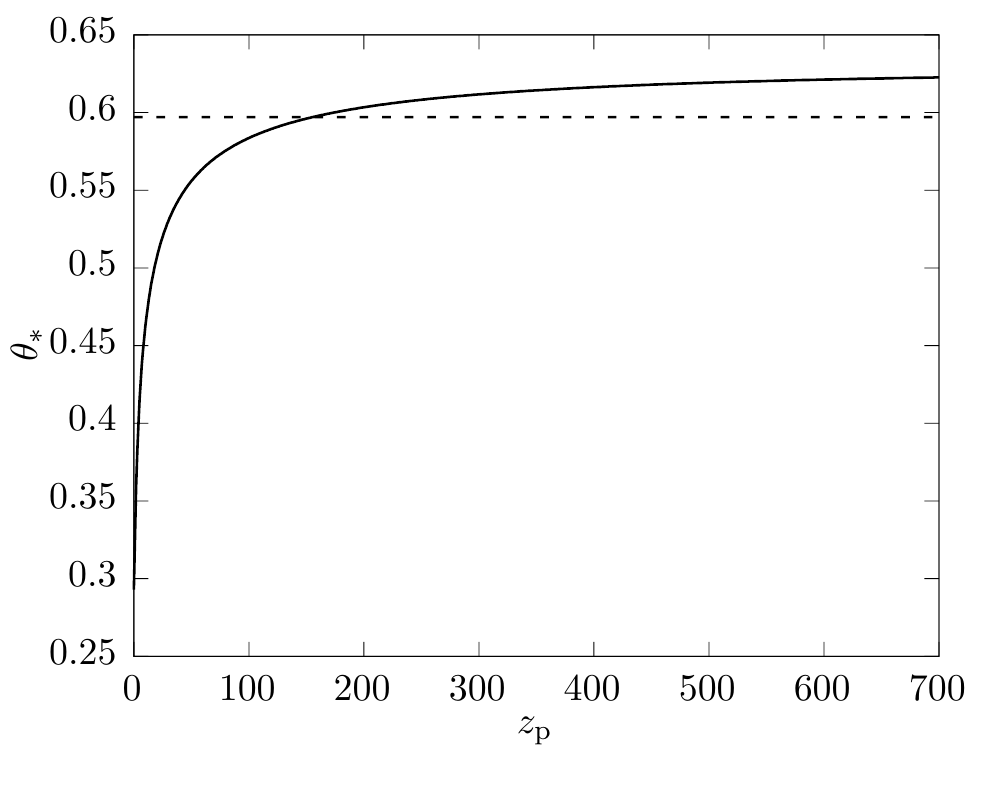}
\caption{Function $\theta_*(z_p)$ for $\Omega_{\Lambda}=0.7$, $\Omega_{{\rm DM},0}=0.26$, 
$\Omega_{b,0}=0.04$, $\Omega_{\gamma,0}=4.6\times 10^{-5}$, and 
$H_0=73.24$\,km\,s$^{-1}$\,Mpc$^{-1}$ for the high-$z$ SU(2)$_{\rm CMB}$ and low-$z$ $\Lambda$CDM interpolating cosmological model considered. Also indicated is the value $\theta_*=0.597^\circ$ (dashed line), fitted to the CMB TT power spectrum.} 
\label{Fig-D1}    
\end{figure}

In order to fix the value of $z_p$ we confront the model of Eqs.\,(\ref{Haxion2}) and (\ref{DSed}) with the observed angular scale $\theta_*$ of the sound horizon at CMB photon decoupling, occurring at $z_{\rm lf,*}$. 
Theoretically, $\theta_*$ is given as
\begin{equation}
\label{angscale}
\theta_*=\frac{r_s(z_{{\rm lf},*})}{\int_0^{z_{{\rm lf},*}}\frac{dz}{H(z)}}\,. 
\end{equation}
To match $\theta_*=0.597^\circ$, as extracted in \cite{Ade2016} from the TT power spectrum, we require $z_p=155.4$, see Fig.~\ref{Fig-D1}. 
This yields a percentage of vacuum energy at CMB photon decoupling of about 
\eqb
\label{vaceneatPD}
\frac{\Omega_{{\rm DM},0}}{\Omega_{b,0}}\left(\frac{z_p+1}{z_{{\rm lf},*}+1}\right)^3\sim 0.65\%\,.
\eqe
The omission of vacuum energy in our SU(2)$_{\rm CMB}$ high-$z$ cosmological 
model of Eq.\,(\ref{Hzexp}) thus is justified for the interpolating model defined in Eqs.\,(\ref{Haxion2}) and (\ref{DSed}).

\section{Summary and outlook}

In the present work we have analysed, based on a modified temperature-redshift relation for the CMB which, in turn, derives from the postulate that thermal photon gases are subject to an SU(2) rather than a U(1) gauge principle, a high-$z$ cosmological model which is void of dark matter and considers three species of massless neutrinos. 
Such a model predicts (after a reconsideration of baryon-velocity freeze-out) a value of the sound horizon $r_s$ which, together with a model independent extraction of the $r_s$--$H_0$ relation from cosmologically local observations in \cite{Bernal2016}, yields good agreement with the value of $H_0$ determined by low-$z$ observations in \cite{Riess2016}. The same $r_s$--$H_0$ relation predicts a low value of $H_0$ in standard $\Lambda$CDM cosmology which is at a 5$\sigma$ discrepancy with the value given in \cite{Riess2016}. 

Motivated by the above results, an interpolation between $\Lambda$CDM at low $z$ and our new high-$z$ model is called for. 
In a first attempt, we have investigated whether coherent and damped oscillations of a Planck-scale axion condensate can realistically accomplish this -- with a negative result. With \cite{HH2017} we were thus led to propose an interpolation in terms of percolated PSA vortices which, at some intermediate $z_p$, partially undergo a depercolation transition. We have demonstrated this model to be consistent with the angular scale of the 
sound horizon at photon decoupling. 

The new model needs to be tested against the various CMB angular spectra. Our hope is that radiative corrections in SU(2) Yang-Mills thermodynamics, which play out at low $z$, are capable of explaining the large-anomalies of the CMB \cite{HofmannNature}.

\end{document}